\documentclass[a4paper]{jpconf}
\usepackage{graphicx}
\begin{document}
\title{Holographic Dual of the Reissner-Nordstr\"om Black Hole}

\author{Chiang-Mei Chen}
\address{Department of Physics and Center for Mathematics and Theoretical Physics, National Central University, Chungli 320, Taiwan}
\ead{cmchen@phy.ncu.edu.tw}

\author{Jia-Rui Sun}
\address{Department of Physics, National Central University, Chungli 320, Taiwan}
\ead{jrsun@phy.ncu.edu.tw}

\begin{abstract}
In this article we give a brief, nevertheless, comprehensive review
on recent studies in the quantum gravity description of the
Reissner-Nordstr\"om (RN) black hole from the perspective of the
AdS/CFT correspondence. We survey the known evidence supporting a
two-dimensional conformal field theory (CFT) description
holographically dual to the RN black hole.
\end{abstract}

\section{Introduction}
The holographic principle~\cite{'tHooft:1993gx, Susskind:1994vu}
provides a novel visual angle to deal with problems of quantum
gravity. It allows us to study the quantum gravity without
quantizing the gravity itself directly, but instead, to find a lower
dimensional equivalently dual field theory which contains the same
information as that of the gravity. However, until now there are
only few explicit realizations of the holographic principle which
have been constructed, for examples the
AdS$_5$/CFT$_4$~\cite{Maldacena:1997re, Gubser:1998bc,
Witten:1998qj} and the AdS$_3$/CFT$_2$~\cite{Brown:1986nw}
correspondences. Recent progress made along this direction was
initiated from the study of the holographic description of the
extremal Kerr black hole---the Kerr/CFT
correspondence~\cite{Guica:2008mu}, which showed that the near
horizon extremal Kerr (NHEK) black hole is dual to the chiral part
of a 2D CFT with central charge $c_L = 12 J$ and temperature $T_L =
1/(2\pi)$, where $J$ is the angular momentum of the Kerr black hole.
This work was later generalized to the near-extremal and nonextremal
Kerr black holes by analyzing the scattering amplitudes of probe
fields both from the gravity and the dual CFT
sides~\cite{Bredberg:2009pv, Castro:2010fd}. The NHEK geometry
containing an AdS$_2$ factor together with an $S^1$ bundle belongs
to the warped AdS$_3$ geometry, thus the Kerr/CFT correspondence is
a kind of the warped AdS$_3$/CFT$_2$ duality. For extension to other
spacetimes which contain a warped AdS$_3$ structure, see the recent
review~\cite{Bredberg:2011hp} and the references therein.

The method in the Kerr/CFT correspondence, however, cannot be directly
applied to non-rotating charged black holes, e.g., the 4D Reissner-Nordstr\"om (RN) black hole, which only contains an apparent AdS$_2$ geometry in the extremal limit, while the $S^1$ bundle is nontrivially encoded in the electromagnetic potential. To study its corresponding CFT dual, so far there are two possible approaches. One is to embed the RN black hole into a higher dimensional spacetime, where part of the background $U(1)$ gauge field forms an extra $S^1$ circle which acts like a kind of rotation in higher dimensional spacetime, and the related analysis can be classified as the Kerr/CFT type, i.e. the warped AdS$_3$/CFT$_2$ correspondence~\cite{Hartman:2008pb, Garousi:2009zx, Chen:2010bs}. Another way, following the Kerr case studied in~\cite{Castro:2009jf}, is to reduce the 4D extremal RN black hole into 2D and then to analyze its 2D effective action based on the boundary counterterm method 
in the AdS$_2$/CFT$_1$ correspondence~\cite{Chen:2009ht}.
These two approaches reveal different sections of the dual CFT: the former one for the left-moving sector and the latter one for the right-moving sector. This interpretation was further confirmed by the studies of CFT dual for the nonextremal RN black hole based on the fact that there exists a hidden 2D conformal symmetry in the background geometry which can be probed by a scalar field at low frequencies~\cite{Chen:2010as, Chen:2010yu}.

To what extent shall we believe in this CFT description for black
holes? As we have mentioned in the beginning, the holographic
principle tells us the dual field theory should contain the same
information as that of the gravity. That means the generating
functionals of the two theories are equivalent. In other words,
besides the matching of entropies of the black hole and the CFT, the
$n$-point correlation functions need also match with each other.
However, until now only 2-point (absorption cross section of
scattering) and 3-point~\cite{Becker:2010jj, Becker:2010dm}
correlation functions have been checked. The matching of higher
point correlators are expected but it requires further study.

This paper is organized as follows. In section II, we give a brief introduction to the 4D RN black hole and its 5D uplifted counterpart, and then in section III, we describe the CFT dual of the (near-)extremal RN black hole in the warped AdS$_3$/CFT$_2$ and in the AdS$_2$/CFT$_1$ correspondences, respectively. In section IV, we move to the CFT dual for nonextremal RN black hole, by probing the hidden conformal symmetry. Finally, we come to the summary and discussion.

\section{Review of the RN black hole}

\subsection{4D RN black hole}
The 4D Reissner-Nordstr\"om solution for a charged black hole is
\begin{eqnarray}\label{4drn}
ds^2 &=& - \left( 1 - \frac{2 M}{r} + \frac{Q^2}{r^2} \right) dt^2 + \frac{dr^2}{1 - \frac{2 M}{r} + \frac{Q^2}{r^2}} + r^2 d\Omega_2^2,
\nonumber\\
A_{[1]} &=& \frac{Q}{r} \, dt, \qquad F_{[2]} = \frac{Q}{r^2} \, dt \wedge dr,
\end{eqnarray}
where $M$ and $Q$ are the mass and electric charge of the black
hole.\footnote{Note that the parameters $M$ and $Q$ have dimension
of length, the corresponding ADM mass and the electric charge
(dimensionless) are $\mathcal{M} = M/G_4, \mathcal{Q} =
Q/\sqrt{G_4}$.} It is the uniquely spherically symmetric
electro-vacuum solution to the equations of motion (Einstein and
Maxwell) of the 4D Einstein-Maxwell theory
\begin{equation}\label{EMaction}
I_4 = \frac1{16 \pi G_4} \int d^4x \sqrt{-g_4} \left( R_4 - F_{[2]}^2 \right).
\end{equation}
The RN black hole is a thermodynamical system with the corresponding Hawking temperature and Bekenstein-Hawking entropy
\begin{equation}\label{TS}
T_H =  \frac{r_+ - r_-}{4 \pi r_+^2}, \qquad S_{BH} =  \frac{\pi r_+^2}{G_4},
\end{equation}
where $r_\pm = M \pm \sqrt{M^2 - Q^2}$ are radii of outer and inner horizons.

The near horizon geometry of the near-extremal RN black hole is~\cite{Bertotti:1959pf, Robinson:1959ev, Maldacena:1998uz}
\begin{eqnarray}\label{nhrn}
ds^2 &=& - \frac{\rho^2 - \rho_0^2}{Q^2} d\tau^2 + \frac{Q^2}{\rho^2
- \rho_0^2} d\rho^2 + Q^2 d\Omega_2^2,
\nonumber\\
A_{[1]} &=& - \frac{\rho}{Q} \, d\tau, \qquad F_{[2]} = \frac1{Q} \, d\tau \wedge d\rho,
\end{eqnarray}
where $\rho_0$ is a small parameter which labels the derivation from extremality and $\rho_0 = 0$ is the extremal limit. The above geometry
is AdS$_2 \times S^2$, and $\rho = \rho_0$ is the horizon, while $\rho \to \infty$ is the AdS$_2$ boundary. In the extremal limit, $T_H = 0$ and $S_{BH} = \pi Q^2/G_4$.

\subsection{5D uplifted RN black hole}
The 4D RN black hole can be uplifted into a 5D solution. Let's consider the 5D Einstein-Maxwell theory
\begin{equation}
I_5 = \frac1{16\pi G_5} \int d^5x \sqrt{- \hat g} \left( \hat R - \frac1{12} \hat F_{[3]}^2 \right),
\end{equation}
and then make the Kaluza-Klein (KK) reduction of the following particular assumption (with a consistent truncation of a constant KK scalar)
\begin{equation}
ds_5^2 = (dy + \mathcal{A}_\mu dx^\mu)^2 + ds_4^2, \qquad \hat A_{[2]} = \hat A_{[1]} \wedge dy,
\end{equation}
where the extra dimension is a circle of radius $\ell$, i.e. $y \sim y + 2 \pi \ell$ and $G_5 = 2 \pi \ell G_4$. Consequently the reduced 4D effective action recovers~(\ref{EMaction})
with the identifications
\begin{equation}
4 F_{[2]}^2 = \mathcal{F}_{[2]}^2 + \hat F_{[2]}^2 = 4 \mathcal{F}_{[2]}^2, \qquad \Rightarrow \qquad \mathcal{A}_{[1]} = A_{[1]}, \quad \hat A_{[1]} = \sqrt3 A_{[1]}.
\end{equation}
Therefore, the 5D uplifted counterpart of the RN solution~(\ref{4drn}) is (it is more convenient to use the ``normalized'' coordinate $\chi = y/\ell$ which has the usual period $\chi \sim \chi + 2\pi$)
\begin{eqnarray}\label{RN5metric}
ds^2 &=& - f(r) dt^2 + \frac{dr^2}{f(r)} + r^2 d\Omega_2^2 + \left( \ell d\chi - \frac{Q}{r} dt \right)^2,
\nonumber\\
\hat A_{[2]} &=& - \sqrt3 \, \frac{Q}{r} \ell \, dt \wedge d\chi.
\end{eqnarray}

The near horizon near extremal geometry is~\cite{Chen:2010bs}
\begin{eqnarray} \label{NHRN5}
ds^2 &=& Q^2 \left[ - (\rho^2 - \rho_0^2) d\tau^2 + \frac{d\rho^2}{\rho^2 - \rho_0^2} + d\Omega_2^2 + \frac{\ell^2}{Q^2} \left( d\hat\chi + \frac{Q}{\ell} \rho d\tau \right)^2 \right],
\nonumber\\
\hat A_{[2]} &=& \sqrt3 \, Q \ell \rho \, d\tau \wedge d\hat\chi,
\end{eqnarray}
where $\hat\chi = \chi - t/\ell$. Note that the size of the extra dimensional circle $\ell$  is in principle a free parameter, thus solution (\ref{NHRN5}) contains a warped AdS$_3 \times S^2$ geometry in general (when $\ell = Q$, it becomes AdS$_3 \times S^2$). In the near extremal limit,
\begin{equation}
T_H = \epsilon \frac{\rho_0}{2 \pi Q^2}, \qquad S_{BH} = \frac{A_5}{4 G_5} = \frac{A_4}{4 G_4} = \frac{\pi}{G_4} (Q^2 + 2 \epsilon Q \rho_0).
\end{equation}
The symbol $\epsilon$ is the limiting parameter for taking the extremal limit, namely $\epsilon \to 0$ corresponds to the extremal case.

\section{(Near-)Extremal RN/CFT}
In this section, we will review the CFT description for the (near-)extremal 4D RN and 5D uplifted RN black holes. Here we will only introduce the basic idea and list the main results, for detailed calculation, please refer to~\cite{Chen:2009ht, Hartman:2008pb, Garousi:2009zx, Chen:2010bs}.

\subsection{Warped AdS$_3$/CFT$_2$ correspondence}
After embedding the 4D RN black hole into 5D, the near horizon extremal geometry~(\ref{NHRN5}) contains the warped AdS$_3$ structure, thus the studying of its dual CFT$_2$ description is rather straightforward.
One can use the techniques in the Kerr/CFT correspondence to calculate the left hand central charge\footnote{For the NHEK geometry, there is another boundary condition discussed in~\cite{Matsuo:2009sj} to compute the right hand central charge directly from the 4D solution. Recently, another attempt was proposed in~\cite{Azeyanagi:2011zj} to calculate both the left and right hand central charges of NHEK at the same time.} and the left hand temperature
\begin{equation}
c_L = \frac{6 Q^3}{G_4 \ell}, \qquad T_L = \frac{\ell}{2 \pi Q}.
\end{equation}
For generality, it is convenient to keep the parameter $\ell$, while the authors in~\cite{Hartman:2008pb, Garousi:2009zx} simply chose a particular value of $\ell = 1$ and the other typical choice $\ell = Q$ has been adopted in~\cite{Chen:2010bs}. Using the Cardy formula, one can check that the CFT entropy reproduces the correct black hole entropy
\begin{equation}
S_{CFT} = \frac{\pi^2}{3} c_L T_L = \frac{\pi Q^2}{G_4}.
\end{equation}

We can further check this warped AdS$_3$/CFT$_2$ correspondence of
the 4D RN black hole at the level of the 2-point correlation
function in the near-extremal limit~\cite{Chen:2010bs}. From the
gravity side this is simply the absorption cross section of probe
fields. For a massless and neutral scalar field of the form
\begin{equation} \label{APhi5}
\Phi(t, r, \theta, \phi, \chi) = \mathrm{e}^{-i \omega t + i n \phi + i k \chi} S(\theta) R(r),
\end{equation}
the solution of the angular part is just the spherical harmonics $Y_{ln}(\theta,\phi)$, and the absorption cross section can be computed for a particular limit $\omega - k/\ell \simeq 0$ as
\begin{equation} \label{sigma}
\sigma_\mathrm{abs} =
\left|\frac{\mathcal{F}^\mathrm{(in)}(r_H)}{\mathcal{F}^\mathrm{(in)}(\infty)}\right|
\sim z_0^{2\beta} \sinh(2 \pi b)  \left| \Gamma\left( \frac12 +
\beta - i a \right) \right|^2 \left|\Gamma\left( \frac12 + \beta - i
(2b - a) \right) \right|^2,
\end{equation}
where the parameters are defined as
\begin{eqnarray}
&& z_0 = \frac{r_+ - r_-}{r_+} = \epsilon \frac{2\rho_0}{Q} =
2\epsilon T_R/T_L,
\nonumber\\
&& a = Q (2\omega - k/\ell) \simeq Q k/\ell, \qquad b = \frac12 \omega Q + \frac{Q (\omega - k/\ell)}{z_0},
\nonumber\\
&& \beta = \sqrt{(l + 1/2)^2 - 6 Q^2 \omega (\omega - k/\ell)} \simeq l + 1/2,
\end{eqnarray}
and $\mathcal{F}^\mathrm{(in)}(r_H)$ is the ingoing flux at the
horizon, while $\mathcal{F}^\mathrm{(in)}(\infty)$ is the ingoing
flux at spatial infinity. Together with the left and right hand
temperatures of the dual 2D CFT $T_L=\frac{\ell}{2 \pi Q}$ and
$T_R=\frac{\ell \rho_0}{2 \pi Q^2}$.

From the CFT side, the absorption cross section corresponds to the 2-point function of the operator dual to the bulk probed scalar field
\begin{equation}\label{CFTp}
P_\mathrm{abs} \sim T_L^{2h_L - 1} T_R^{2h_R - 1} \left( \mathrm{e}^{\pi (\tilde\omega_L + \tilde\omega_R)} \pm \mathrm{e}^{- \pi (\tilde\omega_L + \tilde\omega_R)} \right) \left| \Gamma(h_L + i \tilde\omega_L) \right|^2 \, \left| \Gamma(h_R + i \tilde\omega_R) \right|^2,
\end{equation}
where the conformal weight of the dual operator is $h_L = h_R = \frac12 + \beta$ and
\begin{equation}
\tilde\omega_L = \frac{\omega_L - q_L \mu_L}{2\pi T_L}, \qquad \tilde\omega_R = \frac{\omega_R - q_R \mu_R}{2\pi T_R}.
\end{equation}
The absorption cross sections~(\ref{sigma}) and~(\ref{CFTp}) match with each other by identifying the parameters $q_L = q_R = 0 = \mu_L = \mu_R$ and
\begin{eqnarray}
\tilde\omega_L = a \quad &\Rightarrow& \quad \omega_L \simeq k,
\\
\tilde\omega_R = 2 b - a \quad &\Rightarrow& \quad \omega_R \simeq 2\ell (\omega - k/\ell) \to 0.
\end{eqnarray}

\subsection{AdS$_2$/CFT$_1$ correspondence}
The action~(\ref{EMaction}) of the 4D near horizon extremal RN black hole can be reduced to the 2D effective one
\begin{equation}\label{2Daction}
I_2 = \frac{Q^2}{4 G_4} \int d^2x \sqrt{-g_2} \left( \mathrm{e}^{-2\psi} R_2 + \frac2{Q^2} + 2 \nabla_\mu \mathrm{e}^{-\psi} \nabla^\mu \mathrm{e}^{-\psi} - \mathrm{e}^{-2\psi} F_{[2]}^2 \right),
\end{equation}
by the following ansatz
\begin{equation}
ds^2 = g_{\mu\nu} dx^\mu dx^\nu + Q^2 \mathrm{e}^{-2\psi} d\Omega_2^2, \qquad A = A_\mu dx^\mu,
\end{equation}
where $R_2$ is the 2D Ricci scalar associated with $g_{\mu\nu}$,
$\psi$ is the dilaton field and $F_{\mu\nu} = \partial_\mu A_{\nu} -
\partial_\nu A_{\mu}$ is the $U(1)$ gauge field strength. The 2D
effective action~(\ref{2Daction}) is a classical gravitational
theory which needs a regularization by adding suitable boundary
counterterms. This process is equivalent to making a renormalization
in the CFT side to ensure normalized stress tensor and current. For
the precise implementation, it is convenient to write the 2D
asymptotically AdS$_2$ solution in terms of the Gauss normal
coordinate
\begin{equation}\label{2dmetric}
ds^2 = g_{\mu\nu} dx^{\mu} dx^{\nu} = \mathrm{e}^{-2\psi} dr^2 + \gamma_{tt}(t,r) dt^2, \qquad A = A_t(t,r) dt,
\end{equation}
where $r \to \infty$ is the AdS$_2$ boundary. In this coordinates, the general solution includes a free function of time $f(t)$ (for extremal RN $f = 0$)
\begin{equation}
\gamma_{tt} = - \frac{Q^2}{16} \left( \mathrm{e}^{r/Q} - f(t) \mathrm{e}^{-r/Q} \right)^2, \qquad A_t = - \frac{Q}4 \mathrm{e}^{r/Q} \left( 1 - \sqrt{f(t)} \mathrm{e}^{-r/Q} \right)^2.
\end{equation}

The normalized 2D effective action is
\begin{equation}
I_\mathrm{normalized} = I_2 + I_\mathrm{GH} + I_\mathrm{counter},
\end{equation}
where $I_\mathrm{GH}$ is the Gibbons-Hawking term and $I_\mathrm{counter}$ is the boundary counterterm whose explicit form can be determined by a well-defined variational principle. The normalized boundary stress tensor and currents are defined by
\begin{eqnarray}\label{current}
T_{ab} = - \frac{2}{\sqrt{-\gamma}} \frac{\delta I}{\delta \gamma^{ab}}, \qquad J^a = - \frac1{\sqrt{-\gamma}} \frac{\delta I}{\delta A_a}.
\end{eqnarray}
The right hand central charge of the dual 1D CFT can be obtained from the variation of $T_{ab}$ with respect to a suitable combination of the diffeomorphism and the $U(1)$ gauge transformations~\cite{Chen:2009ht}, and the result is
\begin{eqnarray}
\delta_{\epsilon + \Lambda} T_{tt} &=&  - \frac{Q}{2 G_4} \mathrm{e}^{-\psi} ( \delta_\epsilon \gamma_{tt} + 2 A_t \delta_{\epsilon + \Lambda} A_t )
\nonumber\\
&=& 2 T_{tt} \partial_t \xi + \xi \partial_t T_{tt}
\nonumber\\
&& + \frac{Q^3}{G_4} \left[ \mathrm{e}^{-3\psi} \partial_t^3 \xi - \frac18 \mathrm{e}^{-\psi} \sqrt{f} \, \mathrm{e}^{r/Q} \left( 1 + \sqrt{f} \, \mathrm{e}^{-r/Q} \right) \partial_t \xi \right] \left( 1 - \sqrt{f} \, \mathrm{e}^{-r/Q} \right).
\end{eqnarray}
Comparing with the expected transformation for 1D CFT~\cite{Castro:2008ms}
\begin{equation}
\delta_{\epsilon + \Lambda} T_{tt} = 2 T_{tt} \partial_t \xi + \xi \partial_t T_{tt} - \frac{c}{12} L \partial_t^3 \xi,
\end{equation}
where a dimension of length parameter $L$ is needed to ensure the
central charge to be dimensionless. Actually, this parameter $L$
cannot be fixed only from the 2D effective theory. It is associated
with the radius $\ell$ of the ``internal'' circle to accommodate the
background $U(1)$ gauge field; a reasonable choice is $L = - 2
\ell$. Then the central charge is read out to be
\begin{equation}
c_R = \frac{6 Q^3}{G_4 \ell}.
\end{equation}
Since there is no gravitational anomaly, similar to the near extremal Kerr black hole case~\cite{Castro:2009jf}, $c_L$ should equal to $c_R$. In addition, the left hand temperature of the dual CFT can be calculated from probing the 2D conformal symmetry of the near horizon extremal RN black hole background by a charged scalar field, which gives $T_L = \ell/(2 \pi Q)$. Then using the Cardy formula,
\begin{equation}
S_\mathrm{CFT} = \frac{\pi^2}{3} c_L T_L = \frac{\pi Q^2}{G_4},
\end{equation}
the correct black hole entropy is reproduced.

\section{Nonextremal RN/CFT}
Having discussed the dual CFT description of the (near-)extremal RN
black hole, it is natural to ask what is the holographic dual
description for the general nonextremal RN black hole. However, this
is not an easy problem since there are no apparent AdS structures in
the near horizon geometry of nonextremal RN black holes. The
background conformal symmetry, if it still exists, has been hidden.
Inspired from the experience of studying the scattering process of
probe fields in the black hole background, it was found that a 2D
hidden conformal symmetry actually can be probed by a massless
neutral scalar field propagating in the nonextremal Kerr black hole
background in the low frequency limit~\cite{Castro:2010fd}. In the
following, we will show how this hidden conformal symmetry can be
extracted from the generic nonextremal RN black hole.

\subsection{CFT for 5D uplifted RN black hole}
Like the extremal case, probing of the hidden conformal symmetry in
the 5D uplifted RN black hole background~(\ref{RN5metric}) is
similar to that of the 4D Kerr black hole background. Consider a
bulk massless scalar field $\Phi$ propagating in the background
of~(\ref{RN5metric}), after making the field decomposition
of~(\ref{APhi5}),
the radial part of the Klein-Gordon (KG) equation can be simplified into~\cite{Chen:2010as}
\begin{equation} \label{NHrEQ}
\partial_r (\Delta \partial_r R) + \Biggl[ \frac{\left[ (2 M r_+ - Q^2) \omega - (Q r_+/\ell) k \right]^2}{(r - r_+)(r_+ - r_-)} - \frac{\left[ (2 M r_- - Q^2) \omega - (Q r_-/\ell) k \right]^2}{(r - r_-)(r_+ - r_-)} \Biggr] R = l (l + 1) R.
\end{equation}
where $\Delta = (r - r_+) (r - r_-)$, under the conditions of (a) low frequency and low momentum limit $M \omega \ll 1, \; M k/\ell \ll 1$ (automatically implies $Q \omega \ll 1, \; Q k/\ell \ll 1$), and (b) near region limit of $r \omega \ll 1$ and $r k/\ell \ll 1$. It turns out that the eq.~(\ref{NHrEQ}) is indeed of the same form with the Casimir operator of the $SL(2,R)_L\times SL(2,R)_R$ Lie algebra
\begin{equation}\label{casimir1}
\mathcal{H}^2 = \partial_r (\Delta \partial_r) - \frac{r_+ - r_-}{r - r_+} \left( \frac{T_L \!+\! T_R}{4 \mathcal{A}} \partial_t - \frac{n_L \!+\! n_R}{4 \pi \mathcal{A}} \partial_\chi \right)^2 + \frac{r_+ - r_-}{r - r_-} \left( \frac{T_L \!-\! T_R}{4 \mathcal{A}} \partial_t - \frac{n_L \!-\! n_R}{4 \pi \mathcal{A}} \partial_\chi \right)^2,
\end{equation}
where $\mathcal{A} = n_L T_R - n_R T_L$. Here the eq.~(\ref{casimir1}) is obtained from the standard Casimir operator in the Poincar\'e coordinates ($w^\pm, z$) by the following coordinate transformation
\begin{eqnarray}
w^+ &=& \sqrt{\frac{r - r_+}{r - r_-}} \, \exp(2 \pi T_R \chi + 2 n_+ t),
\nonumber\\
w^- &=& \sqrt{\frac{r - r_+}{r - r_-}} \, \exp(2 \pi T_L \chi + 2 n_- t),
\nonumber\\
z &=& \sqrt{\frac{r_+ - r_-}{r - r_-}} \, \exp\left[ \pi (T_R + T_L)
\chi + (n_+ + n_-) t \right].
\end{eqnarray}
One motivation for making the above coordinate transformation is
that it transforms the Poincar\'e AdS$_3$ spacetime into the
rotating BTZ black hole by simply replacing $r$ with $r^2$.
Comparing eq.~(\ref{NHrEQ}) with eq.~(\ref{casimir1}), one obtains
\begin{equation}
T_R = \frac{(r_+ - r_-) M \ell}{2 \pi Q^3}, \qquad T_L = \frac{(r_+ + r_-) M \ell}{2 \pi Q^3} - \frac{\ell}{2 \pi Q}, \qquad n_\pm = - \frac{r_+ \mp r_-}{4 Q^2},
\end{equation}
and the conformal weights of the operator dual to the bulk scalar field $(h_L, h_R) = (l + 1, l + 1)$.

Even though we can determine the temperatures and conformal weights
of the dual CFT, we still don't know how to calculate the central
charges in the nonextremal case. However, if we simply assume that
the central charges hold the same form as those in the extremal
limit, i.e. $c_L = c_R = 6 Q^3/(G_4 \ell)$, then the Cardy formula
indeed gives the correct black hole entropy
\begin{equation}
S_{CFT} = \frac{\pi^2}3 (c_L T_L + c_R T_R) = \pi (2 M r_+ - Q^2) = \frac{\pi r_+^2}{G_4} = S_{BH}.
\end{equation}

In addition, the absorption cross sections calculated from the gravity side is~\cite{Chen:2010as}
\begin{equation}\label{absgr}
P_\mathrm{abs} \sim T_L^{2h_L - 1} T_R^{2h_R - 1} \sinh\left( \frac{\omega_L}{2 T_L} + \frac{\omega_R}{2 T_R} \right) \left| \Gamma\left( h_L + i \frac{\omega_L}{2 \pi T_L} \right) \right|^2 \, \left| \Gamma\left( h_R + i \frac{\omega_R}{2 \pi T_R} \right) \right|^2,
\end{equation}
where
\begin{equation}
\omega_L = \frac{(2 M^2 - Q^2) (2 M \ell \omega - Q k)}{Q^3}, \qquad \omega_R  = \frac{2 (2 M^2 - Q^2) M \ell \omega - 2 M^2 Q k}{Q^3}.
\end{equation}
Finally, the $P_\mathrm{abs}$ matches with the finite temperature absorption cross section of the operator dual to the scalar field in a 2D CFT.

\subsection{CFT for 4D dyonic RN black hole}
It would be interesting to find a 2D hidden conformal symmetry for
the nonextremal 4D RN black hole, without uplifting it into 5D. This
will give further evidence that the 4D RN black hole is dual to a 2D
CFT. Indeed, it has been shown that the 2D hidden conformal symmetry
does exist in this case; the key point is that the $U(1)$ bundle of
the 4D dyonic RN black hole can be probed by a charged scalar field.
This can be expected since the $U(1)$ symmetry is actually
associated with the gauge symmetry of the background gauge
field~\cite{Chen:2010yu}.

The charged scalar field $\Phi$ with charge $q$ in 4D dyonic RN
black hole background can be decomposed as $\Phi(t, r, \theta, \phi)
= \mathrm{e}^{-i \omega t + i m \phi} S(\theta) R(r)$, then in the
near region $r \omega \ll 1$, together with the low frequency and
small probe charge limit, the radial part of the KG equation becomes

\begin{equation}\label{scalarEQ}
\partial_r (\Delta \partial_r R) + \left[ \frac{r_+^4 (\omega + A_+ q)^2}{(r - r_+)(r_+ - r_-)} - \frac{r_-^4 (\omega + A_- q)^2}{(r - r_-)(r_+ - r_-)} \right] R = l (l + 1) R,
\end{equation}
where $A_\pm = - Q / r_\pm$ are the gauge potentials at the outer/inner horizons. In the dyonic black hole of electric and magnetic charges $Q$ and $P$, the radii of the inner and outer horizons are $r_\pm = M \pm \sqrt{M^2 - Q^2 - P^2}$.

To compare the eq.~(\ref{scalarEQ}) with the Casimir
operator~(\ref{casimir1}), we should introduce an operator
$\partial_\chi$, which can be naturally considered to act on the
``internal'' $U(1)$ symmetry space of the probe electrically charged
scalar field as $\partial_\chi \Phi = i \ell q \Phi$. Then the
temperatures of the dual CFT are
\begin{eqnarray}
T_L = \frac{(r_+^2 + r_-^2) \ell}{4 \pi Q r_+ r_-}, \qquad T_R = \frac{(r_+^2 - r_-^2) \ell}{4 \pi Q r_+ r_-}.
\end{eqnarray}
Again, one assume the central charges are identical to those of the extremal 4D RN black hole, namely,
\begin{equation}
c_L = c_R = \frac{6 Q (Q^2 + P^2)}{G_4 \ell},
\end{equation}
then one again gets
\begin{equation}
S_{CFT} = \frac{\pi^2}3 ( c_L T_L + c_R T_R ) = \frac{\pi r_+^2}{G_4} = S_{BH}.
\end{equation}

The RN/CFT$_2$ correspondence can be further checked by studying the scattering of the probe charged scalar field in the RN black hole background. We will see that the absorption cross section of the scalar field is in agreement with the two point function of its corresponding operators in the dual 2D CFT.

The absorption cross section calculated from the gravity side is of
the same form as eq.~(\ref{absgr})~\cite{Chen:2010yu}. We can also
check the real-time correlators. The two-point retarded correlator
obtained from the gravity side is simply
\begin{eqnarray}\label{GreenR}
G_R &\sim& \sin\left(\pi h_L + i \frac{\tilde\omega_L}{2 T_L} \right) \sin\left(\pi h_R + i \frac{\tilde\omega_R}{2 T_R} \right)
\nonumber\\
&& \Gamma\left( h_L - i \frac{\tilde\omega_L}{2 \pi T_L} \right) \Gamma\left( h_L + i \frac{\tilde\omega_L}{2 \pi T_L} \right) \Gamma\left( h_R - i \frac{\tilde\omega_R}{2 \pi T_R} \right) \Gamma\left( h_R + i \frac{\tilde\omega_R}{2 \pi T_R} \right),
\end{eqnarray}
with $h_L = h_R = l + 1$, the explicit form of other parameters can be found in \cite{Chen:2010yu}.

From the 2D CFT side, the retarded Green function $G_R(\omega_L, \omega_R)$ can be calculated by an analytic continuation from the Euclidean correlator
\begin{eqnarray}
G_E(\omega_{EL}, \omega_{ER}) &\sim& T_L^{2h_L -1} T_R^{2h_R - 1} \mathrm{e}^{i \frac{\tilde\omega_{EL}}{2 T_L}} \mathrm{e}^{i \frac{\tilde\omega_{ER}}{2 T_R}}
\nonumber\\
&& \Gamma\left( h_L \!-\! \frac{\tilde\omega_{EL}}{2 \pi T_L} \right) \Gamma\left( h_L \!+\! \frac{\tilde\omega_{EL}}{2 \pi T_L} \right) \Gamma\left( h_R \!-\! \frac{\tilde\omega_{ER}}{2 \pi T_R} \right) \Gamma\left( h_R \!+\! \frac{\tilde\omega_{ER}}{2 \pi T_R} \right),
\end{eqnarray}
(where $\omega_{EL} = i \omega_L$, $\omega_{ER} = i \omega_R$, $\tilde\omega_{EL} = \omega_{EL} - i q_L \mu_L$ and $\tilde\omega_{ER} = \omega_{ER} - i q_R \mu_R$) by
\begin{equation}
G_E(\omega_{EL}, \omega_{ER}) = G_R(i \omega_L, i \omega_R), \qquad \omega_{EL}, \; \omega_{ER} > 0,
\end{equation}
and the Euclidean frequencies $\omega_{EL}$ and $\omega_{ER}$ should take discrete values of the Matsubara frequencies at finite temperature
\begin{equation}
\omega_{EL} = 2 \pi m_L T_L, \qquad \omega_{ER} = 2 \pi m_R T_R,
\end{equation}
in which $m_L, m_R$ are integers for bosons and half integers for fermions. At these frequencies, we can see that the retarded Green function matches well with the gravity side computation~(\ref{GreenR}) up to a normalization factor depending on the charges $q_L$ and $q_R$, i.e. the electric charge of the probe scalar field.

\section{Summary and discussion}
In this short article, we give a brief review of recent progress in
the holographic description of the RN black hole, for all the
extremal, near extremal and nonextremal cases. We show that, the
warped AdS$_3$/CFT$_2$ and the AdS$_2$/CFT$_1$ descriptions are
consistent, which means that the 4D RN black hole is dual to a 2D
CFT, and indicates that 1D CFT is actually the chiral part of a 2D
CFT. This RN/CFT$_2$ duality is supported from the matching of
entropies, the absorption cross sections and the real-time
correlators calculated both from the gravity and the dual 2D CFT
side. Nevertheless, there are still some unsolved problems such as
why does the method of probing hidden conformal symmetry work? Can
we calculate the central charges of the dual CFT in the nonextremal
limit? Do the $n$-point correlators match with each other? Further
clarification of these important issues is desirable.

\section*{Acknowledgement}
This work was supported by the National Science Council of the
R.O.C. under the grant NSC 99-2112-M-008-005-MY3 and in part by the
National Center of Theoretical Sciences (NCTS).

\section*{References}

\end{document}